\documentclass[pra,twocolumn,superscriptaddress,floatfix]{revtex4-1}
\usepackage{amsmath,amsfonts,amssymb,amsthm,graphics,graphicx,epsfig,times,bbm}
\usepackage[colorlinks=true,citecolor=blue,linkcolor=blue]{hyperref}
\usepackage[usenames]{color}
\usepackage{subfigure}
\usepackage{dcolumn}
\usepackage{graphicx}
\usepackage{bm}
\usepackage{color}
\usepackage{verbatim}
\usepackage{epstopdf}
\usepackage{amstext}
\usepackage{latexsym}
\usepackage{hyperref}
\usepackage{amsfonts}
\usepackage{psfrag}
\usepackage{xcolor}

\begin{document}

\title{Quantum Memristors in Quantum Photonics}

\author{M. Sanz}
\email{mikel.sanz@ehu.eus}
\affiliation{Department of Physical Chemistry, University of the Basque Country UPV/EHU, Apartado 644, E-48080 Bilbao, Spain}
\author{L. Lamata}
\affiliation{Department of Physical Chemistry, University of the Basque Country UPV/EHU, Apartado 644, E-48080 Bilbao, Spain}
\author{ E. Solano}
\affiliation{Department of Physical Chemistry, University of the Basque Country UPV/EHU, Apartado 644, E-48080 Bilbao, Spain}
\affiliation{IKERBASQUE, Basque Foundation for Science, Mar\'{i}a D\'{i}az de Haro 3, E-48013 Bilbao, Spain}
\affiliation{Department of Physics, Shanghai University, 200444 Shanghai, China}

\begin{abstract}
We propose a method to build quantum memristors in quantum photonic platforms. We firstly design an effective beam splitter, which is tunable in real-time, by means of a Mach-Zehnder-type array with two equal 50:50 beam splitters and a tunable retarder, which allows us to control its reflectivity. Then, we show that this tunable beam splitter, when equipped with weak measurements and classical feedback, behaves as a quantum memristor. Indeed, in order to prove its quantumness, we show how to codify quantum information in the coherent beams. Moreover, we estimate the memory capability of the quantum memristor. Finally, we show the feasibility of the proposed setup in integrated quantum photonics. 
\end{abstract}

\maketitle

Circuit elements whose dynamics intrinsically depends on their past evolution~\cite{Ch71, pieee97/1717, nano24/255201} promise to induce a novel approach in information processing and neuromorphic computing~\cite{UMM,Sc11,Le10,YSS13} due to their passive storage capabilities. These history-dependent circuit elements can be roughly classified as purely dissipative, such as memristors, or both dissipative and non-dissipative, such as memcapacitors and meminductors~\cite{nano24/255201,SPN16}. A classical memristor is a resistor whose resistance depends on the record of electrical charges which crossed through it~\cite{Ch71}. The information about the electrical history is contained in the physical configuration of the memristor, summarized in its internal state variable $\mu$, such that in the (voltage-controlled) memristor $I$-$V$-relationship appears as~\cite{nano24/255201},
\begin{subequations} \label{eq:clmemristor}
\begin{eqnarray}
I(t)&=&F(\mu(t), V(t)) V(t), \label{eq:memrresp}\\
\dot \mu(t) &=& G(\mu(t),V(t)), \label{eq:memrstate}
\end{eqnarray}
\end{subequations}
where the state-variable dynamics, encoded in the real-valued function $G(\mu(t), V(t))$, and the state-variable-dependent conductance function $F(\mu(t)) > 0$, lead to a characteristic pinched hysteresis loop of a memristor when a periodic driving is applied~\cite{CK76}. Memristor technology is currently a promising paradigm to replace in the medium term computing architectures based on transistors for certain specific tasks, because of the low energy consumption~\cite{Ch13}. Indeed, they are energetically more efficient~\cite{JKKCH16}, and the presence of memory seems to make them more powerful for key machine learning tasks, such as image recognition~\cite{Ch13,WL95,GKLvHW96,S-GMPIL-B13,BVKISP15}.

The quantization of these devices with memory, especially the memristor, is a complicated challenge which has only recently been achieved~\cite{PEdVSS16,JDdVSE17,SPN16}. The difficulty lies on the fact that it is necessary to engineer an open quantum system whose classical limit corresponds to the general dynamics given by Eqs.~\eqref{eq:clmemristor}. This question was addressed in Ref.~\cite{PEdVSS16} by replacing the memristor by a tunable resistor, a weak-measurement protocol and classical feedback acting on the system-resistor coupling. It was proven that this system behaves in the classical limit following Eqs.~\eqref{eq:clmemristor}. Additionally, it was proven that the dynamics of this composed system is genuinely quantum and, therefore, might be used for quantum information tasks. Afterwards, an implementation of quantum memristors in superconducting circuits was proposed~\cite{JDdVSE17}, making use of the fact that a memristive behavior emerges naturally in the presence of Josephson junctions~\cite{Jo74}. 

\begin{figure}[t!]
\centering
\includegraphics[width=0.35\textwidth]{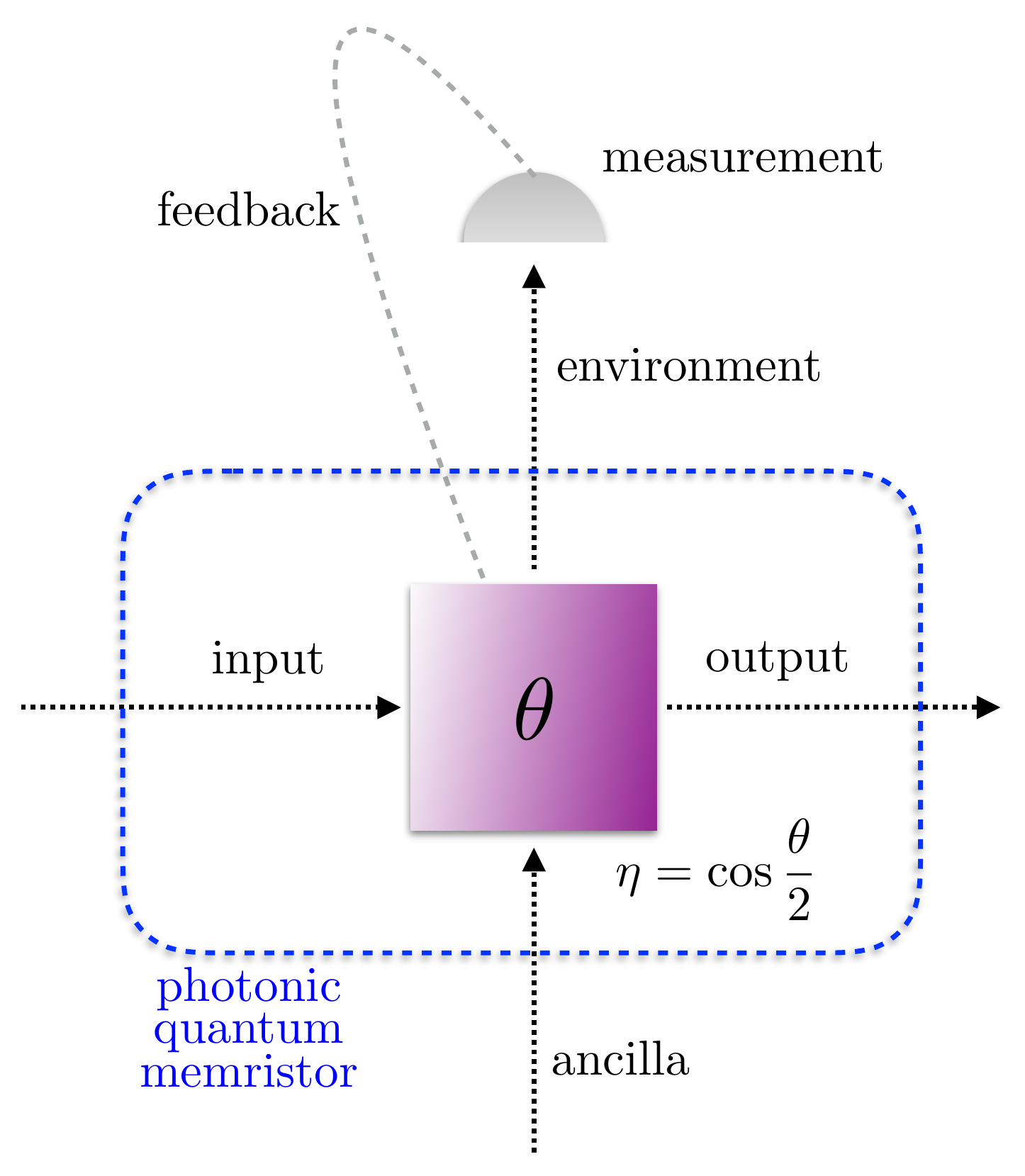}{\vspace{0.5em}}
\caption{Scheme for a quantum memristor in an optical setup by means of a tunable beam splitter, weak measurements and classical feedback.}\label{QM}
\end{figure}

However, a memristive behavior can be extended to a more general framework beyond charges and fluxes. Indeed, a memristive behavior is characterized by non-Markovian history-dependent dynamics, which produces the characteristic pinched hysteresis loops when observables of the input and output states are depicted~\cite{PdV11,JW09}. Additionally, we call these devices quantum memristors when the dynamics cannot be described for all states by means of a classical process. They can be used as building blocks for the simulation of complex non-Markovian quantum dynamics or for an efficient (in terms of resources) codification of quantum machine leaning protocols, mimicking the classical case~\cite{Ch13,WL95,GKLvHW96,S-GMPIL-B13,BVKISP15}.

In this Article, we construct the first quantum memristor in quantum photonics, codifying the quantum information into different quantum states of the photons as depicted in Fig.~\ref{QM}. By showing that the fundamental elements which constitute a quantum memristor, namely a tunable dissipative element, weak measurements and classical feedback~\cite{PEdVSS16}, can be straightforwardly constructed in quantum photonics, we study the dynamics of different initial quantum states and demonstrate the presence of prototypical hysteresis loops. We also compute the persistence of the memory of these devices. We expect that our formalism will provide a toolbox for implementing quantum machine learning in quantum photonics

\begin{figure}[t!]
\centering
\includegraphics[width=0.35\textwidth]{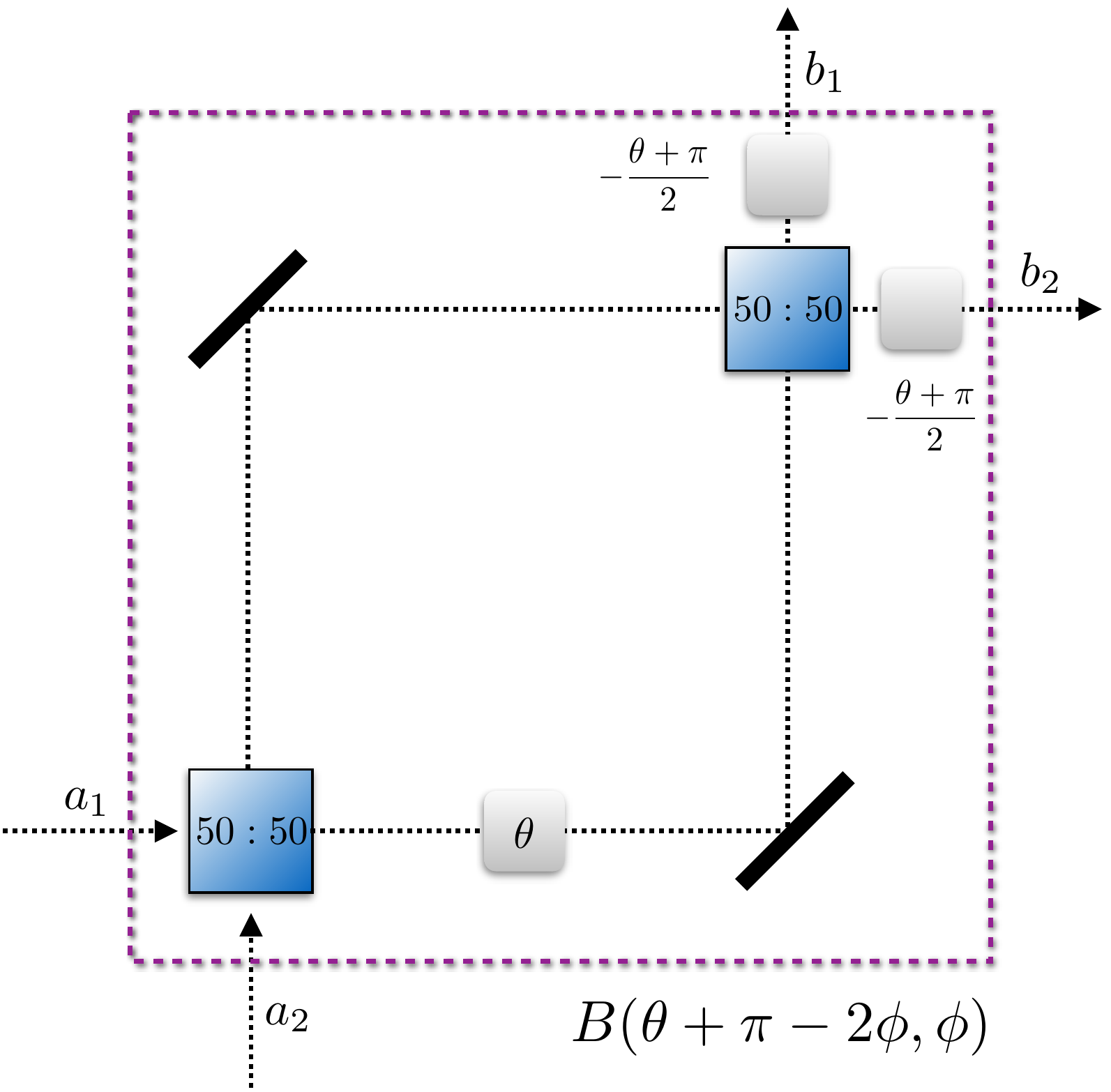}{\vspace{1em}}
 \caption{Two $50 : 50$ beam splitters together with three retarders in a Mach-Zehnder-type array play a role of a beam splitter with arbitrary reflectivity. As the retarder can be straightforwardly in-time controlled, this is suitable for constructing a quantum memristor in quantum photonics.}\label{machzehnder}
\end{figure}

In order to build a photonics quantum memristor, we start by constructing a beam splitter with regulable transmittivity, which will play the role of a tunable coupling to the environment. To achieve it, we use a Mach-Zehnder array with two $50:50$ beam splitters, a retarder, which introduces a phase $\theta$ between arms, and two final compensating phase-shifters (see Fig. \ref{machzehnder}). It can be proven that this construction is equivalent to a beam splitter with an arbitrary reflectivity. Indeed, 
\begin{equation}\label{ArbBS}
B(\Theta,\phi+\frac{\pi}{2}) =e^{-i \frac{\theta}{2}(b_1^{\dagger} b_1+ b_2^{\dagger}b_2)} B(\frac{\pi}{2}, \phi) e^{i \theta \tilde{a}^{\dagger}\tilde{a}} B(\frac{\pi}{2},\phi).
\end{equation}
Here, $\Theta=\pi+\theta-2\phi_T$ is the effective reflectivity of the beam splitter operator, defined as $B(\theta,\phi) = \exp \left [ \frac{\theta}{2} (a_1^{\dagger} a_2 e^{i \phi} -a_1 a_2^{\dagger} e^{-i \phi})\right ]$, $a_1$ and $a_2$ are the annihilation operators in paths $1$ and $2$, respectively, and $\tilde{a}$ the annihilation operator in path $1$ after the first $50:50$ beam splitter. Notice that the phase introduced by the retarder can be straightforwardly controlled in time and, thus, this construction is suitable for our purposes. In order to prove this result, we only need the expression of a 50:50 beam-splitter with transmitted and reflected phases $\phi_T$ and $\phi_R$, respectively, so  
\begin{eqnarray*}
&\frac{1}{2}&
\begin{pmatrix}
e^{i \phi_T} & e^{i \phi_R} \\
-e^{-i\phi_R} & e^{-i\phi_T}
\end{pmatrix} 
\begin{pmatrix}
1 & 0 \\
0 & e^{i \theta}
\end{pmatrix} 
\begin{pmatrix}
e^{i \phi_T} & e^{i \phi_R} \\
-e^{-i\phi_R} & e^{-i\phi_T}
\end{pmatrix}  \\
&=& e^{i \frac{\theta}{2}}
\begin{pmatrix}
e^{i \Phi_T} \cos (\frac{\Theta}{2}) & e^{i \Phi_R} \sin (\frac{\Theta}{2}) \\
-e^{-i \Phi_R} \sin (\frac{\Theta}{2}) & e^{-i\Phi_T} \cos (\frac{\Theta}{2})
\end{pmatrix},
\end{eqnarray*}
which is nothing but $e^{i \frac{\theta}{2}(b_1^{\dagger} b_1+ b_2^{\dagger}b_2)} B(\Theta, \Phi_T, \Phi_R)$, where the parameters are $\Theta = \pi+\theta-2\phi_T$, $\Phi_T = \phi_T+ \frac{\pi}{2}$, and $\Phi_R = \phi_R$, which is an invertible system. The phase difference between the transmitted and reflected phases is $\Phi = \Phi_T - \Phi_R = \phi_T-\phi_R +\frac{\pi}{2} = \phi + \frac{\pi}{2}$, which proves Eq. \eqref{ArbBS}. Now that we can construct an in-time tunable beam splitter, let us study the effect on different initial states, analyzing the hysteretical response and the quantum dynamics of such systems.\\

\begin{figure}[h!]
\centering
\includegraphics[width=0.45\textwidth]{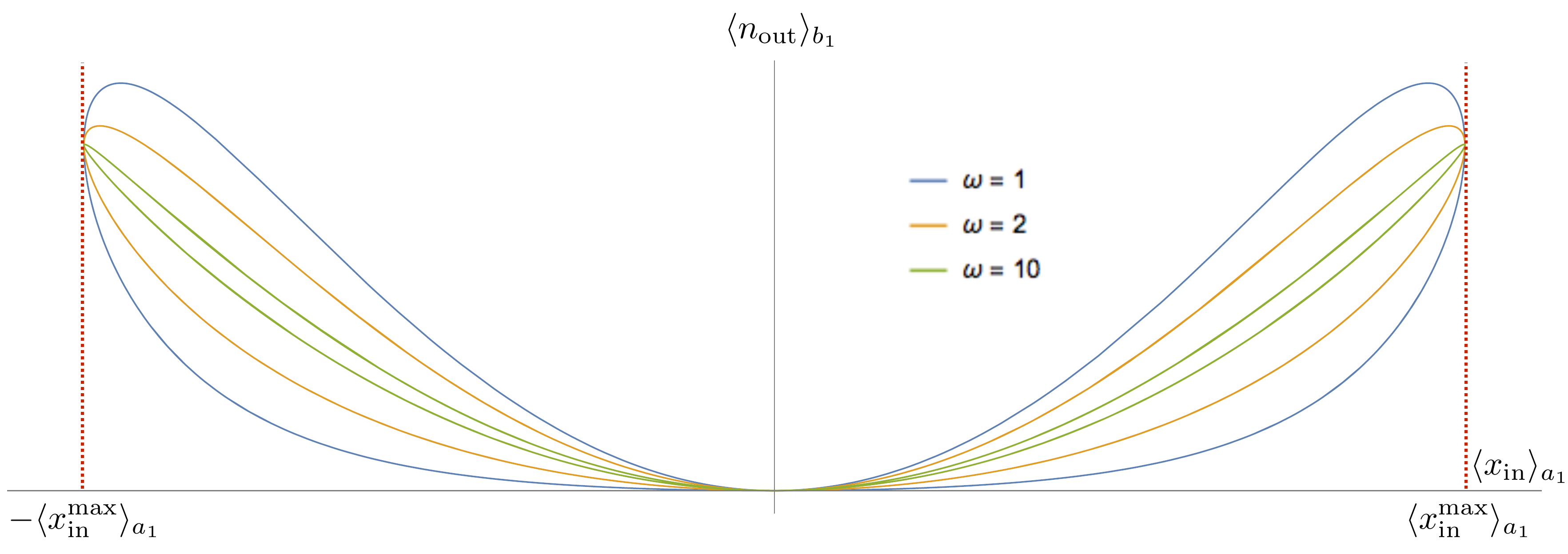}{\vspace{0.5em}}
\caption{Pinched hysteresis loop in the variables $\langle  x_{\text{in}} \rangle_{a_1}$ {\it vs} $\langle \hat{n}_{\text{out}} \rangle_{b_1}$ under a periodic driving. The plot corresponds to three different frequencies of the driving, showing that the enclosed area decreases when the frequency increases. This proves that this system behaves as a memristor.}\label{HL}
\end{figure}

%
% COHERENT STATES
%
{\it Coherent states.} --- Let us start by studying the output of two coherent states after an arbitrary beam splitter \cite{KSBK02}, which is given by $B(\theta,\phi) \mathcal{D}_{a_1}(\alpha)\mathcal{D}_{a_2}(\beta) |0 , 0\rangle = \mathcal{D}_{b_1}(\alpha \cos\frac{\theta}{2} +\beta \sin \frac{\theta}{2} e^{i \phi}) \mathcal{D}_{b_2}(\beta \cos \frac{\theta}{2}-\alpha \sin \frac{\theta}{2} e^{-i\phi}) |0, 0\rangle$. Now, we must prove that, when equipped with measurements and feedback, this system shows a pinched hysteresis behavior, and thus, that it can be a memristor. Let us consider as input states a coherent state $|\alpha\rangle$ in beam $a_1$ and vacuum in beam $a_2$, so the outcome states are $|\alpha \cos \frac{\theta}{2} \rangle_{b_1}$ and $|\alpha \sin\frac{\theta}{2}\rangle_{b_2}$. For the sake of simplicity, we have assumed that $\phi =\pi$, something which can be achieved with an additional phase shifter in the outcome of beam $b_2$. As we are working with coherent states, we will consider as an independent variable $\langle x_{\text{in}} \rangle_{a_1} = \mathbbm{Re}(\alpha)$ of the input beam $a_1$. Assuming a displacement in the $x$-direction, then $\alpha \in \mathbbm{R}$. A memristor is a dissipative element, so we will consider outcoming photon number in beam $b_1$ as the dependent variable, while the beam $b_2$ can be interpreted as losses into a zero-temperature bath. Provided that there is no entanglement between both paths, we make use of outcome beam $b_2$ to measure $\langle n_{\text{out}} \rangle_{b_2} = |\alpha |^2 \sin^2 \frac{\theta}{2}$ and update the value of reflectivity of the beam splitter using phase $\theta$. On the other hand, $\langle n_{\text{out}} \rangle_{b_1} = |\alpha |^2 \cos^2 \frac{\theta}{2}$. Altogether, we can write the following equations for the memristor,
\begin{subequations} \label{eq:memrcoh}
\begin{eqnarray} 
\langle \hat{n}_{\text{out}} \rangle_{b_1} &=& f(\theta, \langle x_{\text{in}} \rangle_{a_1}) \; \langle x_{\text{in}} \rangle_{a_1} \label{eq:memrcoh1}\\
\dot{\theta} &=& g(\theta,\langle x_{\text{in}} \rangle_{a_1}). \label{eq:memrcoh2}
\end{eqnarray}
\end{subequations}
In our case, we pumped the system with a periodic coherent state fulfilling $\langle  x_{\text{in}} \rangle_{a_1} = \langle  x_{\text{in}}^{\text{max}} \rangle_{a_1} \cos (\omega t)$. As $ \langle  x_{\text{in}}^{\text{max}} \rangle_{a_1} = \alpha \in \mathbbm{R}$, we have that $f(\theta, \langle x_{\text{in}} \rangle_{a_1}) = \langle  x_{\text{in}} \rangle_{a_1}  \cos^2 \frac{\theta}{2}$, which may be interpreted as the transmitted intensity per unit of initial displacement. The function $g(\theta,\langle  x_{\text{in}} \rangle_{a_1})$, which updates the reflectivity of the beam splitter, can be chosen freely. In our case, for illustrative proposes, we will select a linear behavior of $\dot{\theta} = \frac{\omega_0}{x_0}\langle  x_{\text{in}} \rangle_{a_1}$. In Fig. \ref{HL}, we have depicted the resulting hysteresis loop when plotting $\langle \hat{n}_{\text{out}} \rangle_{b_1}$ {\it vs} $\langle  x_{\text{in}} \rangle_{a_1}$. As expected, this is a pinched hysteresis loop with an enclosed area decreasing with the frequency of the driving, which shows that this system behaves as a memristor in these variables. As discussed below, this dynamics is deeply related with refractive optical bistability, which makes use of an optical mechanism to change the refractive index inversely to the intensity of the light source \cite{Bo08}, but the classical feedback in our proposal is flexible and we can change the refractive index arbitrarily. 

The area of the hysteresis loop has been proposed as a natural measure of the persistence of the memory in the memristor~\cite{JDdVSE17}. For the general memristor given by Eqs.~\eqref{eq:clmemristor}, this area is given by 
\begin{equation}\label{area}
A = \int_0^T dt \frac{V^2}{2}[\partial_{\gamma} G(\gamma,V,t) W(\gamma,V,t)+ \partial_t G(\gamma,V,t)],
\end{equation}
where the first term in the integrand corresponds to the response related to the selected memory variable and the second term accounts for any remaining explicit time dependence of the conductance. Applying this result to Eqs.~\eqref{eq:memrcoh}, we obtain that the area is given by $A = \frac{\pi \langle  x_{\text{in}}^{\text{max}} \rangle_{a_1}^2}{2 x_0} \frac{\omega}{\omega_0} J_2 \left ( \frac{\langle  x_{\text{in}}^{\text{max}} \rangle_{a_1} \omega_0}{x_0 \omega}\right)$, where $J_2(x)$ is the Bessel function of second order. This formula is valid for $\omega \geq \frac{\langle  x_{\text{in}}^{\text{max}} \rangle_{a_1} \omega_0}{x_0 \pi}$, since the curve suffers additional crosses for smaller frequencies and it must be computed more carefully (in fact, if $\omega < \frac{\langle  x_{\text{in}}^{\text{max}} \rangle_{a_1} \omega_0}{n x_0 \pi}$, $0<n\in\mathbbm{N}$, each bubble suffers $n$ crossings, generating $n+1$ sub-loops). For large frequencies, this area decreases polynomially as $A \sim \frac{\pi \langle  x_{\text{in}}^{\text{max}} \rangle_{a_1}^4 \omega_0}{x_0 \omega}+ \mathcal{O}(\omega^{-3})$, so the area decays slowly and memristive behavior is resilient for a large window of frequencies.

The problem with coherent states is that a beam splitter cannot change the entanglement degree of any state codified in them~\cite{KSBK02}, so the dynamics is essentially classical~\cite{SPGWC10,ICS13,MCD13,USLS14,Fetal15,JWdC16}. \\

\begin{figure}[t!]
\centering
\includegraphics[width=0.40\textwidth]{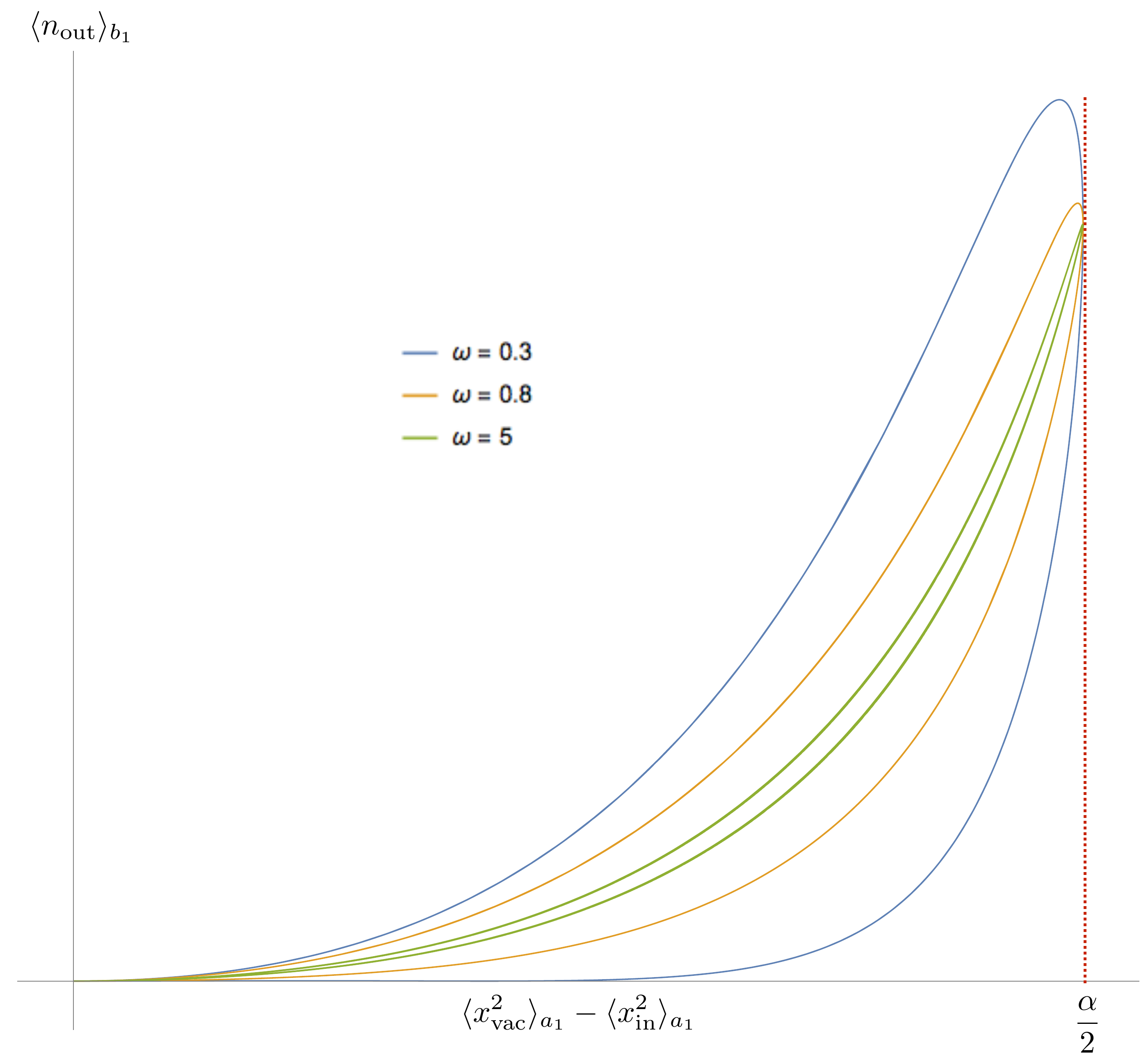}{\vspace{0.5em}}
\caption{Pinched hysteresis loop in the variables $\langle  x^2_{\text{vac}} \rangle_{a_1}-\langle  x^2_{\text{in}} \rangle_{a_1}$ {\it vs} $\langle \hat{n}_{\text{out}} \rangle_{b_1}$ under a periodic driving. The plot corresponds to three different frequencies of the driving, showing that the enclosed area decreases when the frequency increases. For frequencies $\omega < \sqrt{\frac{\alpha}{2}}\frac{\omega_0}{n \pi x_0}$ additional crosses appear generating $n+1$ sub-loops.}\label{HLS}
\end{figure}

%
% SQUEEZED STATES
%
{\it Squeezed states}.--- Let us consider the situation in which the inputs are a squeezed state with squeezing $\zeta = r e^{i \varphi}$ and a vacuum state \cite{BR02}. We can compute the output modes in the Heisenberg picture, which are given for a beam splitter with transmitted and reflective phases $\phi_T$ and $\phi_R$ by
{\small
\begin{eqnarray*}
b_1&=& e^{i \phi_T}\cos \frac{\theta}{2} (a_1 \cosh r  - e^{i \varphi} a_1^{\dagger} \sinh r) + e^{i\phi_R} a_2 \sin \frac{\theta}{2}, \\
b_2 &=& -e^{-i\phi_R} \sin \frac{\theta}{2} (a_1 \cosh r  - e^{i \varphi} a_1^{\dagger} \sinh r) + e^{-i\phi_T}a_2 \cos \frac{\theta}{2}.
\end{eqnarray*}}
In this case, it is straightforward to compute the number of photons in both output beams, so that $\langle n_{\text{out}}\rangle_{b_1} = \sinh^2 r \, \cos^2 \frac{\theta}{2}$ and $\langle n_{\text{out}}\rangle_{b_2} = \sinh^2 r \, \sin^2 \frac{\theta}{2}$. As independent variable, we choose the variance $\langle x^2_{\text{in}}\rangle_{a_1}$, which characterizes a squeezed state and is given by $\langle x^2_{\text{in}}\rangle_{a_1} = \frac{1}{2}(1+ \sinh^2 r -\sinh 2 r \, \cos \varphi)$, where we have taken $\langle x^2_{\text{vac}}\rangle_{a_1}= \frac{1}{2}$ for the vacuum. Hence, the function $f(\theta, \langle x^2_{\text{in}}\rangle_{b_1})$ is 
\begin{equation}
f(\theta, \langle x^2_{\text{in}}\rangle_{b_1}) = \frac{(1-2 \langle x^2_{\text{in}}\rangle_{b_1})^2 \cos^2 \frac{\theta}{2}}{8 \langle x^2_{\text{in}}\rangle^2_{b_1}},
\end{equation}
with $\varphi = 0$, squeezing in the $x$ quadrature. We choose $g(\theta, \langle x^2_{\text{in}}\rangle_{b_1}) = \pm \frac{\omega_0}{x_0 }\sqrt{ \langle x_0^2\rangle-\langle x^2_{\text{in}}\rangle_{b_1}}$, where the sign is chosen depending on the angle, and drive the squeezing below the vacuum variance, so $\langle x^2_{\text{in}}\rangle_{b_1}= \frac{1}{2}(1-\alpha \cos^2 \omega t)$, with $0<\alpha < 1$. Hence, $\theta(t) = \theta_0 + \sqrt{\frac{\alpha \omega_0^2}{2 x_0^2 \omega^2}}\sin \omega t$. The pinched hysteresis loop is depicted in Fig.~\ref{HLS}. 

The area can be computed again by using Eq. \eqref{area}, but in this case, there is no analytical expression. However, the asymptotic expression of the area for large frequencies $\omega\gg 1$ and strong squeezing $\alpha\simeq 1$ can be computed as $A\sim \frac{\pi}{16\sqrt{2}\omega \sqrt{1-\alpha}}$. Therefore, $A$ vanishes in the high-frequency regime as $\mathcal{O}(\omega^{-1})$, which agrees with Fig. \ref{HLS} and also shows resilience of te memory with the frequency.

The dynamics of the beam splitter is quantum when working with squeezed states, and it cannot be simulated by an equivalent classical dynamics. The final state without updating the reflectivity of the beam splitter is given by $\rho_f = \text{tr}_2 [B(\theta,\phi) S_1(\zeta) |00\rangle\langle 00| S_1^{\dagger} (\zeta) B(\theta,\phi)^{\dagger}]$ and the quantum information may be codified in continuous variables~\cite{BvL05}.\\

%
% FOCK STATES
%
{\it Fock states.---} A paradigmatic case of quantum states to codify quantum information corresponds to Fock states. Let us consider a qubit state encoded in a superposition $|\Psi \rangle =e^{i\alpha} \cos \phi \,|0\rangle + \sin \phi\, |1\rangle$ in channel $1$ and the vacuum in input channel $2$, as before. Then, the beam splitter yields $B(\theta, \varphi) |00\rangle = |00\rangle$ and $B(\theta,\varphi)|10\rangle = (\cos \frac{\theta}{2} |10\rangle - e^{i \varphi} \sin \frac{\theta}{2} |01\rangle)$ \cite{KSBK02}. Hence, by linearity, the quantum superposition $|\Psi\rangle$ yields
\begin{eqnarray}\label{eq:finalfock}
B(\theta,\varphi) |\Psi 0\rangle &=& e^{i \alpha} \cos \phi |00\rangle \nonumber \\
&+& \sin \phi \left (\cos \frac{\theta}{2} |10\rangle-e^{-i \varphi}\sin \frac{\theta}{2} |01\rangle \right).
\end{eqnarray}

\begin{figure}[h!]
\centering
\includegraphics[width=0.40\textwidth]{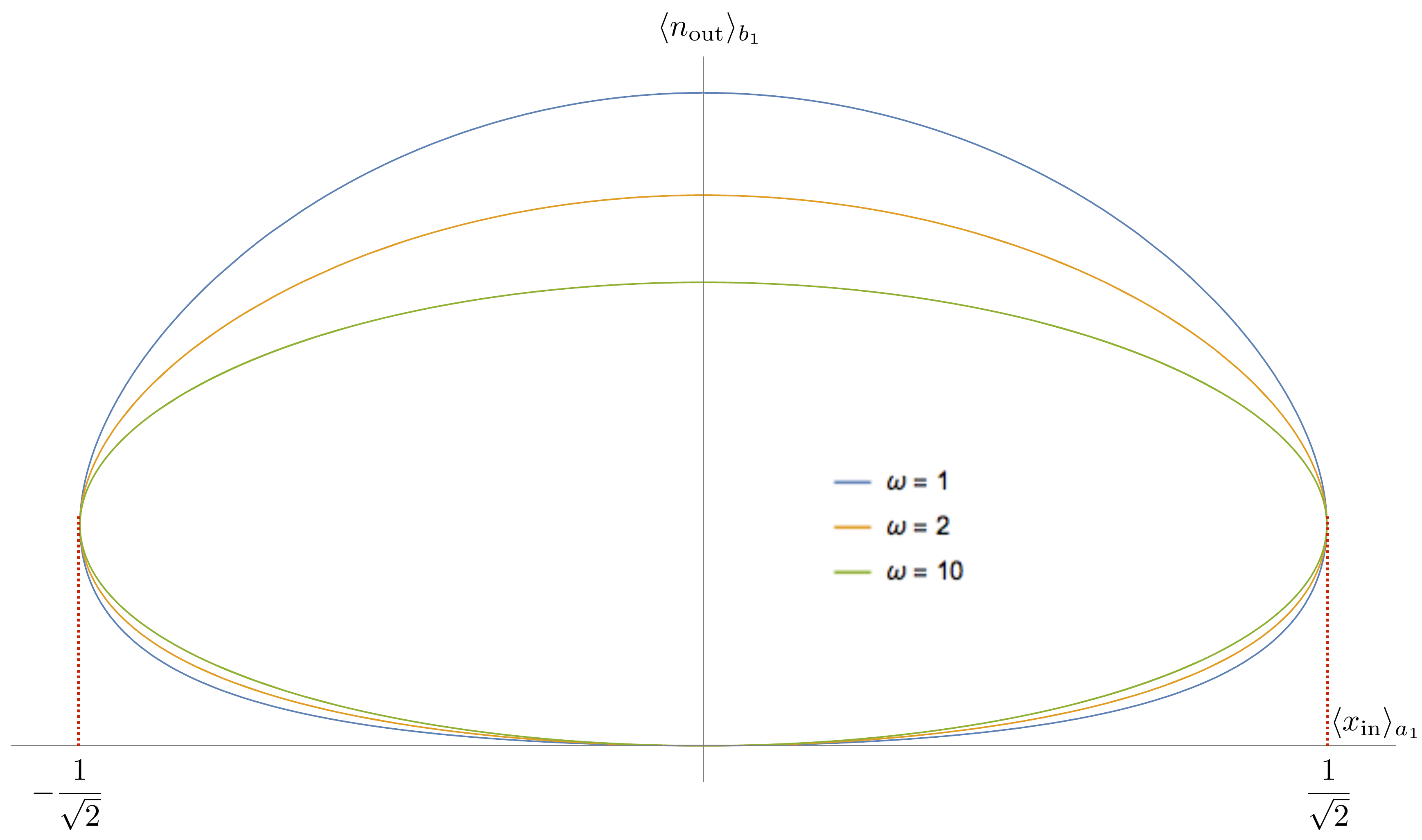}{\vspace{0.5em}}
\caption{Non-pinched hysteresis loop in the variables $\langle \hat{n}_{\text{out}} \rangle_{b_1}$ {\it vs} $\langle  x_{\text{in}} \rangle_{a_1}$ under a periodic driving. The plot corresponds to three different frequencies of the driving, showing that the enclosed area approaches to a constant value $A\sim \frac{\pi}{4\sqrt{2}}$ for high frequencies. For frequencies $\omega < \frac{\omega_0}{(4n-1)\pi}$, $0 < n \in\mathbbm{N}$, $n$ additional crosses of the loop appear.}\label{HLF}
\end{figure}

Let us now prove the memristive behavior of this construction. In order to get it, we have that the mean value of $x$ in the initial state $|\Psi\rangle$ is given by $\langle x_\text{in} \rangle_{a_1} = \frac{1}{\sqrt 2}\sin (2 \phi)$, while the intensity of light coming out through channel $1$ is given by $\langle n_\text{out}\rangle_{b_1} = \sin^2 \phi \sin^2 \frac{\theta}{2}$. Straightforwardly, one obtains that $\langle n_\text{out}\rangle_{b_2} = \sin^2 \phi \cos^2 \frac{\theta}{2}$. Let us consider the same dynamical equation for the internal variable as in the previous cases $\dot \theta  = g(\theta, \langle x_\text{in}\rangle_{a_1}) = \sqrt{2}\omega_0 \langle x_\text{in} \rangle_{a_1} = \omega_0 \sin 2 \omega t$. Then, the hysteresis loops generated, which are valid for frequencies $\omega > \frac{\omega_0}{(4n-1)\pi}$, $0\le n\in\mathcal{N}$, are depicted in Fig.~\ref{HLF}.  For lower frequencies $n$ crosses appear, which generates $n+1$ sub-loops. Differently to the previous cases, the hysteresis loops are not pinched, which means that the memristor is not passive and the energy introduced only vanishes when the initial state is $|0\rangle$. Additionally, this fact also reflects in the fact that the area asymptotically approaches to a constant for high frequencies $A\sim \frac{\pi}{4\sqrt{2}}+\frac{\pi \omega_0}{8\sqrt{2}\omega}$. This robustness could be a useful resource for quantum information processing.

Let us study the final state in channel $1$ when we measure projectively the number of photons in channel $2$ given by Eq.~\eqref{eq:finalfock}. If we measure Fock state $|1\rangle$, then the final state in channel $1$ is simply $|0\rangle$. Otherwise, if we measure Fock state $|0\rangle$ in channel $2$, the final state in channel $1$ is $\frac{e^{i\alpha}\cos \phi}{\sqrt{1-\sin^2 \phi \sin^2 \frac{\theta}{2}}}|0\rangle + \frac{i \sin \phi e^{i\frac{\theta}{2}} \sin\frac{\theta}{2}}{\sqrt{1-\sin^2 \phi \sin^2 \frac{\theta}{2}}}|1\rangle$, which means that the amplitudes of the initial qubit may be modified by means of the reflectivity of the beam splitter. Therefore, this is a natural candidate to be considered for (digital qubit-based) quantum information processing, by creating multiple copies of the same initial state. A possible relevant application could be the simulation of non-Markovian quantum dynamics by considering quantum feedback and $n$ Fock states in input channel $1$ \cite{GZ04}. The reflectivity of the beam splitter should be updated depending on the environment that we want to simulate. \\

{\it Optical bistability.---} Optical bistability is a property of optical devices to show two resonant states, both stable and dependent on the input state. In other words, two different output intensities are possible for a given input intensity, and we need to know the previous states in order to determine which is the right one (non-Markovianity) \cite{Bo08}. This property is characterized by the presence of a hysteresis loop when the output intensity is plotted versus a periodic input intensity. Particularly, refractive bistability makes use of changes in the refractive index of the optical device depending inversely on the intensity of the source light to produce such hysteresis loops with coherent states. This is exactly what we are producing here with different {\it quantum} input states, in such a way that the dynamics shows optical bistability / memristive behavior, but must be described quantum mechanically due to the presence of entanglement. Consequently, quantum information and processing may be encoded in this process, combined with the intrinsic memory (non-Markovianity) of the quantum memristor. Indeed, our system shows all the main elements required for the quantum memristor described in Ref.~\cite{PEdVSS16}.\\

{\it Implementation in quantum photonics.---} The technology for the implementation of our proposal in photonic quantum technology, as for example integrated quantum photonics, is currently available. Indeed, fully reconfigurable two-qubit gates~\cite{Shadbolt} can be directly applied to produce our basic unit of quantum memristor. Moreover, the ease of fabrication of a wide variety of chip designs will allow for establishing a network of quantum memristors based on photonic-chip technology, which could give rise to scalable neuromorphic quantum computing.

There are important differences between the behavior of quantum memristors in superconducting circuits and in quantum photonics. For instance, if we drive a superconducting quantum memristor with an AC voltage/current source which is always positive, for instance $I(t) = I_0 \sin^2 (\omega t)$, a situation closer to our case in quantum photonics, since the photon intensity is always non-negative, then the behaviors are opposed. Instead of just making half of the loop, as one naively could expect, the area of the loop decreases in time until it collapses to a resistive line. The reason is that the resistance can never decrease without negative currents, so it grows until a saturation point in which the memristor does no longer learn and becomes a resistor.\\

{\it Conclusions.---} By using the fundamental elements for the quantization of a memristor, namely a tunable dissipative environment, weak measurements and classical feedback, we have extended the concept of quantum memristor from superconducting circuits to quantum photonics, showing that all these elements are present in current technology. We have studied the dynamics of this photonics quantum memristor with respect to different paradigmatic initial quantum states, showing the prototypical hysteresis loops, computing the corresponding area, and proving that these dynamics are quantum for squeezed states and Fock states. Finally, we have briefly discussed the implementation in integrated quantum photonics. \\

In a long term vision, we expect that these quantum devices can be use as building blocks for quantum machine learning and neural networks~\cite{La17,BWPRWL17,P-OBR-GB17,LA-RM-GSS17} and in the simulation of quantum artificial life~\cite{A-RSLS14,A-RSLS16}. The kind of quantum machine learning algorithms for which these building blocks could be useful is diverse, and ranges from nonlinear quantum neural networks, to supervised learning, unsupervised learning, as well as elements of quantum reinforcement learning. They could also be employed as ingredients of quantum artificial living systems. The main appeal of these systems is that they are fully quantum, at the same time as they provide nonlinear behaviour, highly desirable in the complexity of biomimetic systems and classical learning protocols.

We thank G. Gatti, R. Sweke and F. Flamini for fruitful discussions. This work was supported by Spanish MINECO/FEDER FIS2015-69983-P, Ram\'on y Cajal Grant RYC-2012-11391, and Basque Government IT986-16.

\end{document}